\documentclass[10pt]{article}
\usepackage{epsfig,amsfonts,amssymb}
\usepackage{hyperref, caption}
\usepackage{subfigure}
\usepackage{graphicx}
\usepackage{dcolumn}
\usepackage{bm}
\usepackage{slashbox,amsmath}
\usepackage{epsfig}
\usepackage{etoolbox}
\apptocmd{\sloppy}{\hbadness 10000\relax}{}{}
\input epsf.sty
\usepackage{cite}

\usepackage{graphicx}
\begin{document}

\baselineskip 24pt

\begin{center}
{\Large \bf  Thermodynamics of soft wall AdS/QCD at finite chemical potential }

\end{center}

\vskip .6cm
\medskip

\vspace*{4.0ex}

\baselineskip=18pt

\centerline{\large \rm Shobhit Sachan$^\dagger$ and Sanjay Siwach$^*$}

\vspace*{4.0ex}

\centerline{\large \it Banaras Hindu University}
\centerline{\large \it  Varanasi, 221005, India}

\vspace*{1.0ex}
\centerline{\small E-mail:  $^\dagger$shobhitsachan@gmail.com, $^*$sksiwach@hotmail.com}

\vspace*{5.0ex}

\centerline{\bf Abstract} \bigskip
We study the thermodynamics of soft wall model in AdS/QCD framework. The low temperature phase of QCD is described by thermal AdS and high temperature phase by AdS Black hole
solution of five dimensional gravity with negative cosmological constant. The chemical potential
is introduced via the solution of U(1) vector field in the bulk. The difference of action densities
in two phases is studied and the results are compared with hard wall model. The quark number susceptibility is also calculated in both the models.

\vspace*{3.0ex}
\vfill \eject

\baselineskip=18pt
\section{Introduction}
The strongly coupled regime of gauge theories is full of mysterious phenomena, for example, dynamical
symmetry breaking and dimensional transmutation. QCD provides an interesting example of strongly
coupled gauge theory including the matter fields.  At low energies, the theory is strongly coupled
 and shows confinement. At intermediate energies the theory undergoes a confinement to quark-gluon
 plasma phase transition. Our understanding of this phase transition is limited and the only available
 tool is lattice field theory\cite{shuryak,boyanovsky}.

The techniques of gauge-gravity duality (AdS/CFT correspondence)\cite{Malda,Witten,GKP} are
useful to understand strongly coupled dynamics of gauge theories. The duality captures the dynamics of boundary gauge theories
in terms of dynamics of fields in bulk AdS space time and hence the duality is holographic in
nature. The boundary value of bulk fields couple to currents in boundary theory and act as a
source for them. In general the partition function of boundary theory is equal to partition
function of bulk fields in supergravity approximation. The finite temperature behaviour of gauge
theories is understood in terms of AdS-Schwarzschild black hole geometry in the bulk\cite{Witten1,BISY}.

Recently the AdS/CFT duality is adapted to understand the strong coupling dynamics of QCD\cite{EKSS,KKSS,CPH,RP,EEKT,SKD,nakamura,PGLKS,KLNPS,CABB,MPV,MPV1,TB,PS1,CS,KLY,Erlich,ghergheta,KMSTT,zhang,KKTT}. In the simplest phenomenological model, known as hard wall model, an infra-red boundary cut-off is introduced in
AdS space time\cite{EKSS}. The chiral dynamics is captured by a pair of gauge fields in the bulk and chiral
symmetry is broken by the expectation value of a complex scalar field in the bulk. The boundary
conditions for the bulk gauge fields at infra-red boundary give the spectrum of mesons. In soft
wall model infra-red boundary is pushed to infinity and a dilaton field in the bulk is included
in the action\cite{KKSS}.

In this paper, we investigate the phase structure of holographic QCD in soft wall model at finite chemical potential. We also study the hard wall model and  compare our results with soft wall model. The schematic plots of chemical potential versus transition temperature seem to agree with lattice results\cite{FK,FK1}. We have also calculated the susceptibility in both hard wall and soft wall models.
\section{Soft Wall Model}
Let us consider five dimensional gravity action with negative cosmological constant, $\Lambda = -\frac{6}{L^2}$,
\begin{equation}\label{ag}
S_{grav} ~=~ -\frac{1}{2\kappa^2} \int d^5x \sqrt{g}~ e^{-\Phi}\left(\textrm{R}+\frac{12}{L^2}
\right),
\end{equation}
where $\kappa^2 = 8\pi G_5$ and $G_5$ is five dimensional Newton's constant.

The low temperature phase of boundary gauge theory is described by thermal AdS (tAdS) solution of the above action having line element,
\begin{equation}
ds^2~=~\frac{L^2}{z^2}\left(d\tau^2+dz^2+d\vec{x}^2\right),
\end{equation}
where $0<z<\infty $. The solution to equations of motion for dilaton field is taken as, $\Phi=cz^2$ in order to produce the linear Regge trajectories for mesons\cite{KKSS}.

The high temperature phase is described by AdS black hole (AdSBH) with the line element,
\begin{equation}
ds^2~=~\frac{L^2}{z^2}\left(f(z)d\tau^2+\frac{dz^2}{f(z)}+d\vec{x}^2\right)
\end{equation}
where $f(z)=1-(z/z_h)^4$ and the horizon of the black hole is located at $z_h$. The Hawking temperature of the black hole is given by $T=1/\pi z_h$. In our study, we have used AdS radius $L=1$.

Using action of equation (\ref{ag}), the action density for thermal AdS is given by,
\begin{align}\label{ac1}
\bar{S}_g^{tAdS}~&=~\frac{4}{\kappa^2}\int_0^\beta d\tau \int_\epsilon ^\infty dz\frac{e^{-cz^2}}{z^5}\nonumber\\
&=~\frac{\pi z_h}{\kappa^2}\left(\frac{e^{-c\epsilon^2}(1-c\epsilon^2)}{\epsilon^4}-\frac{e^{-c\epsilon^2}(1-c\epsilon^2)}{2z_h^4}-c^2 \mbox{Ei}( {-c} \epsilon^2)+{\cal O}(\epsilon^4)\right)
\end{align}
where $\beta$ is time periodicity of thermal AdS, given by $\beta=\pi z_h \sqrt{f(\epsilon)}$.  We introduce the UV cut-off $\epsilon$ and take the limit $\epsilon\rightarrow 0$ in the end. The integral `$\textrm{Ei}$' is defined as $\textrm{Ei}(x)~=-\int_{-x}^{\infty}e^{-t}/t~dt$ and $\textrm{Ei}(-\infty)=0$. The action density for AdS black hole solution is given by,
\begin{align}\label{ac2}
\bar{S}_g^{AdSBH}~&=~\frac{4}{\kappa^2}\int_0^{\pi z_h} d\tau \int_\epsilon ^{z_h} dz\frac{e^{-cz^2}}{z^5}\nonumber\\
&=~\frac{\pi z_h}{\kappa^2 z_h^4}\Big[
e^{-c z_h^2} \left( c z_h^2 -1 \right)
+  c^2 z_h^4 \mbox{Ei}( {-c} z_h^2)\Big]\nonumber\\
&~+\frac{\pi z_h}{\kappa^2}\Big[
\frac{e^{-c \epsilon^2}}{\epsilon^4} \left( 1-c \epsilon^2 \right)
-  c^2 \mbox{Ei}( {-c} \epsilon^2)\Big]
\end{align}
where we have defined $\bar{S}=S/V_3$ and $V_3$ is three volume ($V_3\equiv \int d^3x$). On subtracting equation (\ref{ac2}) and (\ref{ac1}), the divergent terms cancel out and in the limit $\epsilon\rightarrow 0$, the difference in gravitational action densities for two solutions is given by,
\begin{eqnarray}
\Delta \bar{S}_{g} &=& \lim_{\epsilon \to 0} ( \bar{S}_g^{AdSBH} -  \bar{S}_g^{tAdS})\nonumber\\
&=& \frac{\pi}{\kappa^2 z_h^3} \Big[
e^{-c z_h^2} \left( c z_h^2 -1 \right) +\frac{1}{2}
+  c^2 z_h^4 \mbox{Ei}( {-c} z_h^2)\Big]
\end{eqnarray}
The difference vanishes for $cz_h^2=0.419 ~ \textrm{or} ~T_c=0.492\sqrt{c}$ where $c=(388 MeV)^2$ and one gets a phase transition from AdS spacetime to AdS black hole solution\cite{CPH}.

We are interested in the finite temperature dynamics of QCD at finite matter density in holographic model described above. The matter part of Euclidean action is given by,
\begin{equation} \label{bulk2}
 S_{matter} =  \frac{1}{4g_5^2} \int d^5x ~e^{-\Phi}\sqrt{g}~F^2
\end{equation}
where $g_5$ is the five dimensional gauge coupling constant.

The quark number density is identified with the zero momentum limit of time component of $U(1)$ vector field of the chiral symmetry group $U(2)_L\times U(2)_R$, which obeys,
\begin{equation}
 \partial_z \left( \frac{1}{z} e^{-\Phi} \partial_z V_\tau(z) \right)  = 0.
\end{equation}
The solution of the above equation is given by,
\begin{equation} \label{vec1}
 V_\tau(z)~=~ae^{cz^2}~+~b
\end{equation}
where $a$ and $b$ are integration constants. Our solution diverges for $z\rightarrow\infty$. In order to regularize the solution, the upper limit of the solution is taken as $z_\Lambda$. For thermal AdS solution, $z_\Lambda \rightarrow \infty$ and for AdS black hole solution $z_\Lambda=z_h$. Using boundary conditions, $V_\tau (z=z_\Lambda)=0$ and $V_\tau (z=0)=i\mu$, the general solution can be written as,
\begin{equation}
V_\tau(z)~=~\frac{i \mu}{1-e^{c {z_\Lambda}^2}}(e^{cz^2}-e^{c {z_\Lambda}^2}).
\end{equation}
The parameter $\mu$ is introduced for later convenience and can be identified as quark chemical potential \cite{nakamura}. The factor $i$ is used here because we are dealing with Euclidean signature  \cite{LPS,OA}. The boundary condition $V_\tau (z=z_h)=0$ for AdS black hole solution is needed to ensure the regularity of the solution at the horizon. For thermal AdS this boundary condition $V_\tau (z=z_\Lambda\rightarrow\infty)=0$ is needed for the finiteness of the solution.

Using this, the action density for thermal AdS solution is given by,
\begin{align}
\bar{S}_v^{tAdS}~&=~-\lim_{\epsilon\rightarrow 0}\frac{\mu^2~ c^2}{ g_5^2 (e^{c{z_\Lambda}^2}-1)^2}~\int_0^{\beta} d\tau \int_\epsilon^{z_\Lambda} dz ~ze^{cz^2}\nonumber\\
&=~-\frac{\mu^2~c}{g_5^2} ~\pi z_h~\frac{1}{(e^{c{z_\Lambda}^2}-1)}.
\end{align}
In the limit $z_\Lambda\rightarrow\infty$, we get $\bar{S}^{tAdS}_v=0$, i.e. no contribution in grand potential from thermal AdS side.

Similarly, the action density for AdS black hole is given by,
\begin{equation}
\bar{S}_v^{AdSBH}~=~-\frac{\mu^2~c}{g_5^2} ~\pi z_h~\frac{1}{(e^{cz_h^2}-1)}.
\end{equation}

The difference in action densities for vector part of matter action densities in both solutions is given by,
\begin{equation}
\Delta \bar{S_v}~=~-\frac{\mu^2~c}{g_5^2} ~\pi z_h~\frac{1}{(e^{cz_h^2}-1)}.
\end{equation}

Grand potential $\Omega(\mu , T)$ is related with on-shell Euclidean action $S$ as
\begin{equation}
\Omega(\mu ,T) = T S_{\mathrm{on-shell}}
\end{equation}
 where $T$ is the black hole temperature.

 Using $\frac{1}{\kappa^2}~=~\frac{N_c^2}{4\pi^2},~\frac{1}{g_5^2}~=~\frac{N_c N_f}{4\pi^2}$, the difference in grand potentials of AdS black hole and thermal AdS is given by,

\begin{equation}
\frac{\Delta\Omega}{V_3}=\frac{1}{z_h^4}\frac{N_c^2}{4\pi^2}\left[
e^{-cz_h^2} \left(  c z_h^2 -1 \right) +\frac{1}{2}+  c^2 z_h^4 \mbox{Ei}( {-c} z_h^2)\right]
-\frac{c~\mu^2~ N_c~N_f}{4\pi^2~(e^{cz_h^2}-1)}.
\end{equation}
which can be written in terms of black hole temperature $T=\frac{1}{\pi z_h}$,
\begin{equation}\label{der1}
\frac{\Delta\Omega}{V_3} =\frac{\pi^2T^4N_c^2}{4}\left[
e^{-c/\pi^2T^2} \left(\frac{c}{\pi^2T^2} -1 \right) +\frac{1}{2}+  \frac{c^2}{\pi^4T^4}\mbox{Ei}\left(-\frac{c}{\pi^2T^2}\right)\right]
-\frac{c~\mu^2~ N_c~N_f}{4\pi^2~\left(e^{c/\pi^2T^2}-1\right)}.
\end{equation}

The entropy (${\cal S}$) in the grand canonical formalism is defined as ${\cal S}=-(\partial\Omega/\partial T)_{V,\mu}$. The entropy difference between the two phases is given as,
\begin{equation}
\frac{\Delta {\cal S}}{V_3}~=~-\frac{1}{2} e^{-c/\pi ^2 T^2} \left(e^{c/\pi ^2 T^2} - 2\right) \pi ^2 T^3 N_c^2
+\frac{c^2 e^{c/\pi ^2 T^2} \mu ^2 N_c N_f}{2 \left(e^{c/\pi ^2 T^2} - 1\right)^2 \pi ^4 T^3},
\end{equation}
which is a nonzero at transition temperature, indicating a first order phase transition.

The quark number susceptibility is given by,
\begin{equation}
\chi (T)~=~-\frac{1}{V_3}\frac{\partial^2 \Omega^{AdSBH}(\mu ,T)}{\partial\mu^2}
=~\frac{c~ N_c N_f}{2 \pi ^2\left(e^{c/\pi ^2 T^2} - 1\right)}.
\end{equation}
For large values of $T$ ($T>>\sqrt{c}$), we have $e^{c/\pi ^2 T^2}\approx 1+c/\pi ^2 T^2$. In this limit the expression for $\chi$ becomes,
\begin{equation}
\chi(T)~\approx~\frac{N_c N_f T^2}{2},
\end{equation}
which is same as in hard wall model as we shall see later.
From now onwards we shall take $N_f=2$ and $N_c=3$ unless stated otherwise.

\begin{figure}[htp]
\centering
\includegraphics[scale=0.90]{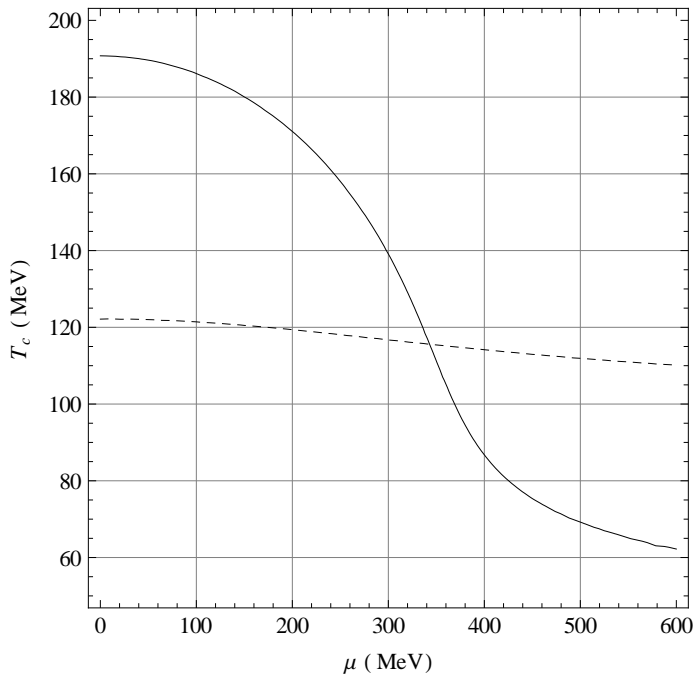}
\caption{Transition temperature vs chemical potential in soft wall model (continuous curve) and in hard wall model (dashed curve).}
\label{Tmu}
\end{figure}

In fig. \ref{Tmu} we have plotted the transition temperature as a function of chemical potential. This is obtained by setting the difference of grand potential in two phases equal to zero. At high temperature the grand potential is lower for AdS black hole, which corresponds to deconfined phase in QCD. As the temperature is lowered, thermal AdS solution becomes favourable, corresponding to confined phase of QCD. The temperature where $\Delta\Omega=0$, we get Hawking -Page transition in supergravity picture. 
\section{Hard Wall Model}
Now let us compare our results with the hard wall model. The action in hard model is given as
\begin{align}
S~&=~S_{grav}+ S_{matter}\nonumber\\
&=~ -\frac{1}{2\kappa^2} \int d^5x \sqrt{g}~ \left(\textrm{R}+\frac{12}{L^2}\right)+\frac{1}{4g_5^2} \int d^5x \sqrt{g}~F^2
\end{align}

In hard wall model, an infrared cut-off is introduced in boundary theory at some finite value of $z=z_{IR}$\cite{EKSS}. The numerical value of $z_{IR}=(323 ~MeV)^{-1}$ is determined by fitting the data for low lying meson states with experiments. The solution of the time component of $U(1)$ vector field is given by,
\begin{equation}
 V_\tau (z)~=~c_1\,+\,c_2\,z^2.
\end{equation}
where the constant $c_1$ is related to quark chemical potential and $c_2$ is related to quark number density as $c_2=12\,\pi^2\,n_q/N_c$ \cite{SKD}.

In hard wall model we use boundary conditions, $V_\tau(z=z_{\Lambda})=0$ and $V_\tau(z=0)=i\mu$. Using these boundary conditions, the general solution can be written as,
\begin{equation}
V_\tau (z)~=~i\mu\left(1-\frac{z^2}{z_{\Lambda}^2} \right).
\end{equation}
For thermal AdS, $z_\Lambda=z_{IR}$ and for AdS black hole, $z_\Lambda=z_h$. For AdS black hole this condition is required to ensure the regularity of the solution at the horizon. Although for thermal AdS, there is no prior reason to impose the boundary condition $z_\Lambda=z_{IR}$, but we use it. This gives a linear relation between quark number density and quark chemical potential, consistent with the relation found in ref. \cite{SKD}. This boundary condition also implies that the chemical potential is zero whenever the charge density vanishes.

After evaluating the action densities of gravitational \cite{CPH} and matter part in both phases, the difference in grand potential of the two phases (AdS black hole and thermal AdS) is given by,
\begin{eqnarray}
\frac{\Delta\Omega}{V_3}&=& \left\{\begin{array}{ll}
\frac{1}{\kappa^2} \frac{1}{2z_h^4}-\frac{\mu^2 }{g_5^2}\left(\frac{z_{IR}^2}{z_h^4}-\frac{1}{z_{IR}^2}\right) &~~~ z_{IR} < z_h\\ \\
\frac{1}{\kappa^2}\left(\frac{1}{z_{IR}^4}-\frac{1}{2z_h^4}\right)-\frac{\mu^2 }{g_5^2}\left(\frac{1}{z_h^2}-\frac{1}{z_{IR}^2}\right)&~~~ z_{IR} > z_h.
\end{array}\right.
\end{eqnarray}
For $z_{IR}<z_h$, the expression is always positive, therefore no phase transition. We study the condition $z_{IR} > z_h$ for phase transition. Putting values $\frac{1}{\kappa^2}~=~\frac{N_c^2}{4\pi^2}~\textrm{and}~\frac{1}{g_5^2}~=~\frac{N_c N_f}{4\pi^2}$, we get difference of grand potential in hard wall model as,
\begin{equation}\label{der2}
\frac{\Delta\Omega}{V_3} = \frac{ N_c^2}{8}\pi^2\left(\frac{2}{\pi^4z_{IR}^4}-T^4\right)-\frac{\mu^2 N_c~N_f}{4}\left(T^2-\frac{1}{\pi^2 z_{IR}^2}\right),
\end{equation}
We have plotted transition temperature $T_c$ vs chemical potential $\mu$ (when $\Delta\Omega =0$) in fig \ref{Tmu}. In our calculations in this model, we have used $z_{IR}=1/323~ MeV^{-1}$.
The entropy difference between the two phases in this model is given by,
\begin{equation}
\frac{\Delta {\cal S}}{V_3}~=~\frac{N_c^2}{2}\pi^2T^3 +\frac{N_c~N_f}{2}\mu^2~T.
\end{equation}
The susceptibility in hard wall model is given by,
\begin{equation}
\chi(T)~=-\frac{1}{V_3 }\frac{\partial^2 \Omega^{AdSBH}(\mu ,T)}{\partial\mu^2}~=~\frac{N_cN_f}{2}T^2.
\end{equation}
which agrees with soft wall model for higher temperatures ($T>>\sqrt{c}$).
\section{Conclusion}
The plot of transition temperature versus chemical potential seems to agree with lattice results \cite{FK,FK1} in both the models. In soft wall model there is more rapid change in transition temperature for values of $\mu$ between 200-400 $MeV$. Since the entropy difference $\Delta{\cal S}\neq 0$ at the transition temperature, the phase transition is of first order. We have also evaluated the  susceptibility in both hard wall and in soft wall models. Our results are valid for small values of the chemical potential as we haven't considered back reacted geometry. Perhaps a more useful setting for back reacted geometry is charged AdS black hole solution as investigated by Lee et. al.\cite{LPS}. It would be interesting to study back reaction of matter on geometry and the transport properties like electrical conductivity etc. for the system.


\section*{Acknowledgement}
The work of S. Sachan is supported by CSIR-Junior Research Fellowship and S. Siwach is
supported by DST, Govt. of India project (No.P-07-472).

\end{document}